\documentclass[aps,pra,twocolumn,groupedaddress,showpacs]{revtex4}

\usepackage{graphicx}
\usepackage{amsmath}
\usepackage{bm}

\begin{document}


\title{Wall-vortex composite solitons in two-component Bose-Einstein condensates}


\author{Kenichi Kasamatsu$^{1}$}
\author{Hiromitsu Takeuchi$^{2}$}
\author{Makoto Tsubota$^{3}$}
\author{Muneto Nitta$^{4}$}
\affiliation{
$^1$Department of Physics, Kinki University, Higashi-Osaka, 577-8502, Japan \\
$^2$Graduate School of Integrated Arts and Sciences, Hiroshima 
University, Kagamiyama 1-7-1, Higashi-Hiroshima 739-8521, Japan \\
$^3$Department of Physics and The Osaka City University Advanced
Research Institute for Natural Science and Technology (OCARINA),
Osaka City University, Sumiyoshi-ku, Osaka 558-8585, Japan \\ 
$^4$Department of Physics, and Research and Education Center for Natural 
Sciences, Keio University, Hiyoshi 4-1-1, Yokohama, Kanagawa 223-8521, Japan }


\date{\today}

\begin{abstract}
We study composite solitons, consisting of domain walls and vortex lines 
attaching to the walls in two-component Bose-Einstein condensates. 
When the total density of two components is homogeneous, the system can be mapped to 
the O(3) nonlinear sigma model for the pseudospin representing the two-component 
order parameter and the analytical solutions of the composite solitons can 
be obtained. Based on the analytical solutions, we discuss the detailed structure of the 
composite solitons in two-component condensates by employing the 
generalized nonlinear sigma model, where all degrees of freedom of the original 
Gross-Pitaevskii theory are active. 
The density inhomogeneity results in reduction of the domain wall tension 
from that in the sigma model limit. 
We find that the domain wall pulled by a vortex is logarithmically bent as a membrane pulled 
by a pin, and it bends more flexibly than not only the domain wall in the sigma model 
but also the expectation from the reduced tension. 
Finally, we study the composite soliton structure for actual experimental situations with trapped immiscible 
condensates under rotation through numerical simulations of the coupled Gross-Pitaevskii equations. 
\end{abstract}


\maketitle
\section{INTRODUCTION}\label{Intro}
Topological defects or topological solitons are solutions of systems obeying partial differential equations, 
representing localized structures with their stability being due to non-trivial topology \cite{Manton}. 
Vortices in superfluids/superconductors are an example of line topological defects 
\cite{Donneley}, and it is believed that the analogous defects would exist in 
early universe as cosmic strings \cite{Kibble}. 
A domain wall is a planer topological defect separating two different vacua or phases.
When a symmetry group $G$ of a system is spontaneously broken to a subgroup $H$, 
topologically allowed defect type is determined by the homotopy properties of the 
order parameter space (vacuum manifold) $G/H$. In a $(d+1)$-dimensional spacetime, $p$-dimensional defects 
($p<d$) exist if the homotopy group $\pi_{d-p-1} (G/H)$ is nontrivial. 
Thus, for $d=3$ there will be planar defects (domain walls) if 
$\pi_{0}(G/H) \neq 0$, linear defects (vortices or strings) if $\pi_{1}(G/H) \neq 0$, 
and point defects (monopoles) if $\pi_{2}(G/H) \neq 0$. 
These defects can be classified as ``singular" or ``continuous" in a sense 
whether (a part of) $G$ is recovered at the core of defects or not. 
Order parameter is not defined at the core of a singular defect, while it is 
defined everywhere for continuous texture (defects). 

Bose-Einstein condensates (BECs) of ultra-cold atomic gases provide 
an ideal system for examining topological solitons in a quantum condensed system \cite{Pethickbook}. 
A major advantage of this system is that the properties of BECs can be quantitatively
described using the mean-field theory, namely, the Gross-Pitaevskii (GP) model. 
From experimental point of view, cold atom BECs are a versatile system 
to study topological defects, because most of the system parameters are tunable 
and optical techniques allow one to engineer the condensate wave function as well as 
to visualize the condensates directly. 
In the context of a single-component BEC characterized by scalar order parameter with the 
broken U(1) symmetry, there are many papers discussing 
the properties of solitons and vortices; see Refs. \cite{Frantzeskakis,Fetterreview} for reviews. 
In addition, realization of multicomponent (spinor) BECs with multiple order parameters 
provides a ground to study more complex topological solitons \cite{Kawaguchirev}, 
as studied in superfluid $^3$He \cite{Volovik}. 
For example, the dark-bright solitons can be excited in two-component 
BECs \cite{Becker,Hamner}, 
where a dark soliton (density dip) of one component can trap a bright soliton 
(density hump) of the other component \cite{Busch}. 
Exotic vortices composed of several order parameter components were 
observed experimentally \cite{Matthews,Leanhardt,Schweikhard,Leslie,Choi}. 
Because the order parameter space of the multicomponent 
BECs can possesses higher symmetry than U(1) of the scalar BEC, their homotopy groups $\pi_n$ with 
different $n$ can become simultaneously nontrivial and thus different kinds of topological 
solitons can coexist. There have been discussed the structure, stability, 
and creation/detection schemes 
for various kinds topological solitons in multicomponent BECs, such as 
monopole \cite{Stoof,Martikainen,Savage,Ruostekoski2,Pietila}, 
three-dimensional (3D) skyrmion \cite{AlKhawaja,Ruostekoski,Battye,Savage2,Herbut,Kawakami}, 
cosmic vortons \cite{Metlitski,Nitta}, and knots \cite{Cho,Kawaguchi}. 

\begin{figure}
\begin{center}
\includegraphics[width=0.98 \linewidth,keepaspectratio]{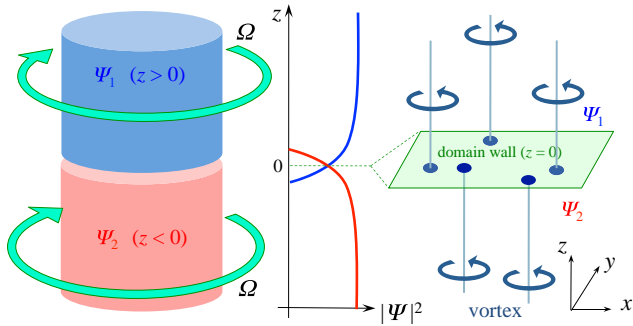}
\end{center}
\caption{(Color online) Schematic illustration of the wall-vortex soliton 
configuration in two-component BECs. 
The two-component BECs $\Psi_{1}$ at $z>0$ and $\Psi_{2}$ at $z<0$ are 
separated by the domain wall at the $z=0$ plane. 
Since penetration of the condensate densities takes place around 
the boundary, the domain wall is well-defined as a plane on which both components 
have the same amplitude, as shown in the middle panel. 
We assume that vortex lines are straight and perpendicular to the wall. 
Rotation is applied to the two components with the different rotation frequency 
${\bf \Omega}_i = \Omega_i \hat{\bf z}$ } 
\label{Fig1ponch}
\end{figure}
In this paper, we discuss a 3D composite soliton  
consisting of domain walls and vortices in immiscible two-component BECs, 
as sketched in Fig.~\ref{Fig1ponch}. 
Two-component BECs have been realized by using the mixture of atoms with two 
hyperfine states of $^{87}$Rb \cite{Myatt,Hall,Mertes,Tojo} 
or the mixture of two different species of atoms such 
as $^{87}$Rb-$^{41}$K \cite{Modugno,Thalhammer}, 
$^{85}$Rb-$^{87}$Rb \cite{Papp} or $^{87}$Rb-$^{133}$Cs \cite{McCarron}. 
The experiments \cite{Tojo,Papp} demonstrated that miscibility 
and immiscibility of two-component BECs can be controlled by tuning the atom-atom 
interaction via Feshbach resonances. 
The domain wall is referred to as a boundary of 
phase-separated two-component BECs and is well-defined as a plane 
on which both components have the same amplitudes \cite{Tim,Ao,Coen,Barankov}. 
The vortices can be arranged by applying rotation of the confining 
potential around the $z$-axis to the phase-separated BECs \cite{Fetterreview}. 
We assume that  two components undergo phase separation 
in the $z>0$ and $z<0$ region, forming a domain wall lying at the $z=0$ plane 
and edges of the vortex lines along the $z$-axis attach to the domain wall. 

In our previous paper \cite{KasamatsuD}, we pointed out that the wall-vortex composite 
soliton in two-component BECs can be identified as a non-relativistic analog of 
``Direchlet (D)-brane soliton" found in some field theoretical 
models \cite{Gauntlett,Shifman,Isozumi,Sakai,Eto:2006pg}. 
This statement is based on the fact that 
the GP equations for two-component BECs can be mapped to 
the O(3) nonlinear sigma model (NL$\sigma$M) by introducing a pseudospin 
representation of the order parameter \cite{KTUreview,Kasamatsu2,Mason}. 
The NL$\sigma$M admits the solitonic object that can have similar 
properties to the D-brane in the string theory \cite{Gauntlett}. 
The purpose of this paper is to discuss in more detail the structure of this 
composite soliton in two-component BECs. 
The generalized NL$\sigma$M for two-component BECs includes additional 
degrees of freedom compared with the original O(3) NL$\sigma$M, 
which modifies some properties of the composite soliton known in the 
previous literatures: (i) Vortices consisting of the composite soliton have singular core, 
while they are nonsingular in the NL$\sigma$M.
(ii) The density inhomogeneity of the BECs results in reduction of domain wall tension 
from that in the NL$\sigma$M. 
(iii) The domain wall attached by a vortex is logarithmically bent, as a membrane pulled 
by a pin, and it bends more flexibly than not only the domain wall in the NL$\sigma$M 
but also the expectation from the reduced tension.
We also study the composite soliton structure of rotating immiscible BECs in a 
trapping potential through 3D numerical simulations of the coupled GP equations. 
To reduce the gradient energy of the density, the domain wall tends to be parallel 
to the rotation axis and forms a vortex sheet \cite{Kasamatsu3}. 
At high rotation frequency, a lattice of 2D skyrmion can form upon the domain wall, 
which undergoes triangular or rectangular ordering caused by the effective 
intercomponent repulsion realized in the restricted system on the domain wall. 

This paper is organized as follows. 
In Sec.~\ref{formulation}, we formulate the problem for two-component BECs and 
introduce the pseudospin representation to reduce the GP model into the NL$\sigma$M. 
In Sec.~\ref{Gauntlettreview}, we examine the structure of wall-vortex composite solitons 
based on the analysis of the NL$\sigma$M, where analytic solutions of these solitons 
can be obtained.  
In Sec.~\ref{numeri}, we discuss how the composite solitons in two-component 
BECs are modified from the analytic solutions in the NL$\sigma$M
and presents the results of 3D numerical simulations 
for the trapped immiscible two-component BECs under rotation. 
We conclude this paper in Sec.~\ref{concle}. 

\section{Theoretical formulation of two-component BECs}\label{formulation}
We study the detailed properties of the composite solitons in two-component BECs, 
whose basic configuration is illustrated schematically in Fig.~\ref{Fig1ponch}. 
Two-component BECs are represented by the order parameters 
$ (\Psi_1, \Psi_2)^T = (\sqrt{\rho_1} e^{i \theta_1} ,  \sqrt{\rho_2} e^{i \theta_2})^T $, 
which are the condensate wave functions with the density $\rho_j$ 
and the phase $\theta_j$ ($j=1,2$). 
They are confined in some trapping potentials and 
undergo phase separation, which results in the domain walls. 
The quantized vortices can exist in each component, being 
created by rotating the system or imprinting the circulating phase by 
the atom-laser coupling \cite{Fetterreview}. 
We first show that the theoretical formulation 
of this system can be mapped to the NL$\sigma$M. 
This mapping was firstly discussed in the two-component 
Ginzburg-Landau theory for charged two-component Bose systems \cite{Babaev}, 
which was applied to the two-component BECs by some of the authors \cite{KTUreview,Kasamatsu2}. 

The solitonic structure in two-component BECs is given by the analysis of the 
two-component GP model. The energy functional is given by
\begin{eqnarray}
E [\Psi_{1},\Psi_{2}] = \int d {\bf r} \biggl\{ \sum_{j = 1,2} 
\biggl[ \frac{\hbar^{2}}{2m_{j}}  \left| \left( \nabla - i \frac{2m_j}{\hbar} \tilde{\bf A} \right) \Psi_{j} \right|^{2} \nonumber \\ 
+ (V_j- \mu_{j}) |\Psi_{j}|^{2} 
+ \frac{g_{jj}}{2} |\Psi_{j}|^{4} \biggr]  
+ g_{12} |\Psi_{1}|^{2} |\Psi_{2}|^{2} \biggr\}.   
\label{energyfunctio2}
\end{eqnarray}
Here, $m_{j}$ and $\mu_j$ are the mass and the chemical potential of the $j$th component, respectively. 
The trapping potential is written by an axisymmetric harmonic oscillator as 
\begin{equation}
V_j = \frac{1}{2} m_j \omega_j^2 (r^2 + \lambda^2 z^2) 
\label{trap}
\end{equation}
with an aspect ratio $\lambda$, where $\lambda <1$ ($> 1$) represents a cigar-shaped 
(pancake-shaped) potential. 
The coefficients $g_{11}$, $g_{22}$, and $g_{12}$ represent the 
atom-atom interactions. They are expressed in terms of the 
$s$-wave scattering lengths $a_{11}$ and $a_{22}$ 
between atoms in the same component and $a_{12}$ between atoms 
in the different components as 
\begin{equation}
g_{jk} = \frac{2 \pi \hbar^{2} a_{jk}}{m_{jk}}
\end{equation} 
with $m_{jk}^{-1} = m_{j}^{-1} + m_{k}^{-1}$. The vector potential $\tilde{\bf A}$ is generated by 
(i) the rotation of the system $\tilde{\bf A} = ({\bf \Omega} \times {\bf r}) / 2 $ \cite{Fetterreview} 
or (ii) a synthesis of the artificial magnetic field by the laser-induced Raman coupling between 
the internal hyperfine states of the atoms \cite{Lin}. 

The two-component GP model can be transformed to the similar form of the NL$\sigma$M 
by introducing the pseudospin representation of the order parameter. 
Here, we confine ourselves to the simple situation with the equal mass 
$m_1=m_2=m$ and equal trapping frequency $\omega_1 = \omega_2 = \omega$; 
its derivation in the case of the 
general parameters of the system, e.g., the mass 
imbalance and the difference of the trapping frequencies, 
was considered by Mason and Aftalion \cite{Mason}. 
The condensate wave functions are denoted as 
\begin{eqnarray} 
\left( 
\begin{array}{c}
\Psi_{1} \\
\Psi_{2}
\end{array} 
\right)
= \sqrt{\rho} e^{i \Theta /2}
\left( 
\begin{array}{c}
 \zeta_1 \\
 \zeta_2
\end{array} 
\right) .
\end{eqnarray}
Here, $\zeta=[\zeta_1,\zeta_2]^T$ is the spin-1/2 spinor with 
$|\zeta_1|^2 + |\zeta_2|^2 = 1$. 
The four degrees of freedom of the original 
wave functions $\Psi_{j} = \sqrt{\rho_{j}}e^{i\theta_{j}}$ 
(their amplitudes $\rho_{j}$ and phases $\theta_{j}$) are expressed 
in terms of the total density $\rho = \rho_{1} + \rho_{2}$, the total 
phase $\Theta=\theta_{1}+\theta_{2}$, 
and the polar angle $\theta$ and azimuthal angle $\phi$ of the local
pseudospin ${\bf s}=(s_{x},s_{y},s_{z})$ defined as 
\begin{eqnarray}
{\bf s}=\zeta^{\dagger}  \bm{\sigma}  \zeta
= \left[ 
\begin{array}{c}
\zeta_{1}^{\ast} \zeta_{2} + \zeta_{2}^{\ast} \zeta_{1} \\
-i(\zeta_{1}^{\ast} \zeta_{2} - \zeta_{2}^{\ast} \zeta_{1})\\
|\zeta_{1}|^{2} - |\zeta_{2}|^{2}
\end{array} 
\right]
= \left[ 
\begin{array}{c}
\sin \theta \cos \phi \\
\sin \theta \sin \phi \\
\cos \theta
\end{array} 
\right] , \label{spindef}
\end{eqnarray}
where ${\bm \sigma}$ is the Pauli matrix, $\cos \theta = (\rho_{1} - \rho_{2})/\rho$, 
$\phi = \theta_{2}-\theta_{1}$, and $|{\bf s}|^2=1$. 
By using these variables, the total energy Eq.~(\ref{energyfunctio2}) can be rewritten as the form of 
the generalized NL$\sigma$M \cite{Kasamatsu2}:
\begin{eqnarray}
E = \int  d {\bf r}  \biggl\{ \frac{\hbar^{2}}{2m} \biggl[ 
(\nabla \sqrt{\rho})^{2}+ \frac{\rho}{4} \sum_{\alpha} 
(\nabla s_{\alpha})^{2} \biggr]  + V \rho \nonumber \\
+ \frac{m \rho}{2} v_{\rm eff}^{2}  
+ U(\rho,s_z) \biggr\}, 
\label{nonsigmamodBEC}
\end{eqnarray}
where we have introduced the effective velocity field ${\bf v}_{\rm eff} = {\bf v}_g + {\bf v}_s - 2 \tilde{\bf A}$ 
coming from the gradient of the total phase: 
\begin{eqnarray}
{\bf v}_g = \frac{\hbar}{2m} \nabla \Theta 
\end{eqnarray}
and the flux flow of the spinor:
\begin{eqnarray}
{\bf v}_s &=& \frac{\hbar}{2mi} \sum_{j=1,2} (\zeta_j^{\ast} \nabla \zeta_j - \zeta_j \nabla \zeta_j^{\ast} ) \nonumber \\
&=& \frac{\hbar}{2m} \frac{s_z}{s_x^2+s_y^2} (s_y \nabla s_x - s_x \nabla s_y ) \nonumber \\ 
&=& - \frac{\hbar}{2m} \cos \theta \nabla \phi. 
\end{eqnarray}
Here, we have also used the relation 
\begin{equation}
\frac{\hbar^2}{2m}  ( |\nabla \zeta_1|^2 + |\nabla \zeta_2|^2 ) - \frac{v_s^2}{4} = \frac{\hbar^2}{8m} \sum_{\alpha} (\nabla s_{\alpha})^2. 
\end{equation}  
The second term in the right hand side of Eq.~(\ref{nonsigmamodBEC}) corresponds to the classical NL$\sigma$M 
for Heisenberg ferromagnet. The generalized NL$\sigma$M has several unique features 
that are revealed as : (i) There is a gradient term of the total density. 
 (ii) The spin stiffness, a prefactor of the $(\nabla s_{\alpha})^{2}$ term, 
is dependent on the total density $\rho$ and is generally spatially inhomogeneous. 
(iii) There is an additional kinetic-energy term $m \rho v_{\rm eff}^{2}/2$, 
associated with the presence of the superfluid velocity ${\bf v}_{\rm eff} \neq 0 $ and the 
external vector potential $\tilde{\bf A} \neq 0$. 

The potential $U$ is a function of the total density $\rho$ and the $z$-component $s_z$ of the pseudospin only, 
being explicitly written as 
\begin{equation}
U(\rho, s_z) = c_{0} + c_{1} s_{z} + c_{2} s_{z}^{2}  \label{becsigmapotential}
\end{equation} 
with 	
\begin{eqnarray}
c_{0} &=& \frac{\rho}{8} [\rho (g_{11}+g_{22}+2g_{12}) 
- 4 (\mu_{1} + \mu_{2}) ],  \\ 
c_{1} &=& \frac{\rho}{4}[ \rho (g_{11}-g_{22}) - 2 (\mu_{1} - \mu_{2}) ],  \\ 
c_{2} &=& \frac{\rho^2}{8} (g_{11}+g_{22}-2g_{12}).
\end{eqnarray} 
If $g_{11} \neq g_{22} \neq g_{12}$ or $\mu_1 \neq \mu_2$, the anisotropic terms with the coefficients $c_{1}$ 
and $c_{2}$ break the global SU(2)-invariance of the system. 
The coefficient $c_{1}$ can be interpreted as a longitudinal magnetic field 
that likes to align the pseudospin along the $z$-axis. 
The term with the coefficient $c_{2}$ determines the spin-spin interaction 
associated with $s_{z}$; it is antiferromagnetic for $c_{2}>0$ and 
ferromagnetic for $c_{2}<0$ \cite{Kasamatsu2}. 
The stationary point of this potential gives the equilibrium values 
\begin{eqnarray}
\rho = \frac{(g_{22}-g_{12}) \mu_1 + (g_{11}-g_{12}) \mu_2 }{g_{11} g_{22} -g_{12}^2}, \\
s_z = \frac{(g_{22}+g_{12}) \mu_1 - (g_{11}+g_{12}) \mu_2 }{(g_{22}-g_{12}) \mu_1 + (g_{11}-g_{12}) \mu_2 } .
\end{eqnarray}
The determinant of the Hessian at that point is given by 
\begin{equation}
\frac{\partial^2 U}{\partial \rho^2} \frac{\partial^2 U}{\partial s_z^2} 
- \left( \frac{\partial^2 U}{\partial s_z \partial \rho} \right)^2 
= \frac{[(g_{22}-g_{12}) \mu_1 + (g_{11}-g_{12}) \mu_2]^2}{4 (g_{11} g_{22} - g_{12}^2)}.
\end{equation}
The stationary point is a minimum or a maximum only when $g_{11} g_{22} - g_{12}^2 > 0$. Otherwise, 
the minimum of the potential disappears within the range $-1<s_z<1$ and the 
degenerate energy minima are given by $s_z=1$ or $s_z = -1$. 
This situation corresponds to the ferro-magnetization, namely, the phase separation 
of the two-component BECs, which is discussed in the following. 


\section{Topological solitons in nonlinear sigma model}\label{Gauntlettreview}
In order to understand the properties of wall-vortex (D-brane) soliton in field theoretical model, 
we review the work by Gauntlett {\it et al}. \cite{Gauntlett}. 
The NL$\sigma$M is a scalar field theory whose (multi-component) scalar field  
defines a map from a `space-time' to a 
Riemann (target) manifold. The massive hyper-K\"{a}hler 
sigma model employed by Gauntlett {\it et al.} corresponds to the massive NL$\sigma$M for 
the effective description of the Heisenberg ferromagnet with spin-orbit coupling. 
The energy functional is given as 
\begin{equation}
E[{\bf s}] = \frac{1}{4} \int d {\bf r} \left[ \sum_{\alpha=1}^{3} (\nabla s_{\alpha})^{2} + U({\bf s}) \right],
\label{nsigmam}
\end{equation}
also known as the Landau-Lifshitz model governing the high-spin and long 
wavelength limit of ferromagnetic materials. 
Here, the amplitude of the vector is $|{\bf s}({\bf r})|=1$ everywhere. 
The ground state is two-fold degenerate such as $s_{z} = +1$ and $-1$, where 
the potential is now described as 
\begin{equation}
U({\bf s}) = m_{\sigma}^{2} (1-s_{z}^{2})
\end{equation}
with a mass parameter $m_{\sigma}$. 

Under several conditions, our model Eq.~(\ref{nonsigmamodBEC}) can be reduced to the same form of 
Eq.~(\ref{nsigmam}) \cite{KasamatsuD}. 
For the simple situation, we consider the homogeneous system without the trapping potential $V=0$ 
and put the parameters as $g_{11} = g_{22} \equiv g$ and $\mu_{1} = \mu_{2} \equiv \mu$. 
The anisotropy coefficient $c_1$ in Eq.~(\ref{becsigmapotential}) then vanishes and 
the total energy can be written as 
\begin{eqnarray}
E = \int  d {\bf r}  \biggl\{ \frac{\hbar^{2}}{2m} \biggl[ 
(\nabla \sqrt{\rho})^{2}+ \frac{\rho}{4} \sum_{\alpha} 
(\nabla s_{\alpha})^{2} \biggr]  
+ \frac{m \rho}{2} v_{\rm eff}^{2}  \nonumber \\  
+  \frac{g}{2} \left( \rho - \frac{\mu}{g} \right)^2  - \frac{g-g_{12}}{4} \rho^2 (1-s_z^2)  \biggr\}, 
\label{becsimplenlsm}
\end{eqnarray}
where the constant term has been omitted. 
The coefficient of the last term in Eq.~(\ref{becsimplenlsm}) is positive 
because we consider the case of $g_{12} > g$, giving the mass for the $s_z$-field.
For the limit $g \to \infty$, which corresponds to the Thomas-Fermi limit \cite{Pethickbook}, 
we can approximate that the total density is frozen to be $\rho = \mu / g \equiv \rho_0$ and 
the $(\nabla \sqrt{\rho})^2$-term vanishes. 
The kinetic energy associated with the superflow ${\bf v}_{\rm eff}$ is assumed to be negligible 
for simplicity \cite{tyuua}. 
By using the healing length $\xi = \hbar/\sqrt{2 m g \rho_0}$ as the length scale, 
the total energy reduces to 
\begin{equation}
\tilde{E} = \frac{E}{g \rho_0^2 \xi^{3}}  \simeq  \frac{1}{4} \int  d {\bf r} \left[ \sum_{\alpha} (\nabla s_{\alpha})^{2} 
+ m_{\sigma}^{2} (1 - s_{z}^{2}) \right] \label{nonsigmamod2}
\end{equation}
with the mass 
\begin{equation}
m_{\sigma}^{2} \equiv \left| 1-\frac{g_{12}}{g} \right|. 
\label{2becnlsmmass}
\end{equation}
Therefore, the following discussion based on Eq.~(\ref{nsigmam}) can be applied 
approximately to our system. Actually, as seen later, the additional degrees of freedom 
in two-component BEC system yield only quantitative modification of the soliton structure. 

\begin{figure}
\begin{center}
\includegraphics[width=0.98 \linewidth,keepaspectratio]{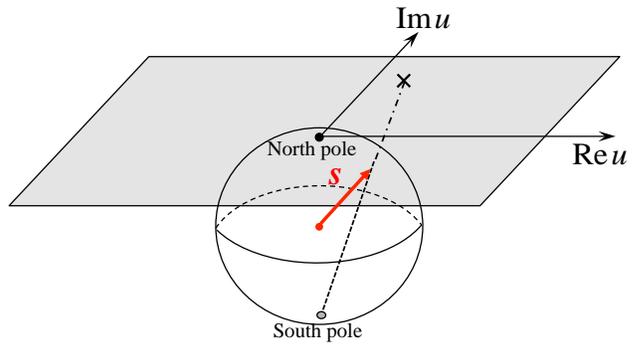}
\end{center}
\caption{(Color online) Stereographic projection from the sphere to the tangent plane at the north pole. } 
\label{stereo}
\end{figure}
To this end, we introduce the stereographic coordinate 
\begin{equation}
u=\frac{s_{x}+is_{y}}{1+s_{z}},
\end{equation}
where $u=0$ $(\infty)$ 
corresponds to the north (south) pole of the spin sphere, as shown in Fig.~\ref{stereo}. 
Then, each component of the pseudospin is written as 
\begin{eqnarray}
\left(s_{x} , s_{y} , s_{z} \right) 
= \left(   \frac{u+u^{\ast}}{1+|u|^{2}}, - i \frac{u-u^{\ast}}{1+|u|^{2}},  \frac{1-|u|^{2}}{1+|u|^{2}} \right) 
\end{eqnarray} 
and Eq.~(\ref{nsigmam}) becomes
\begin{equation}
E = \int d^3 x \frac{\sum_{\alpha} |\partial_{\alpha} u|^{2} + m_{\sigma}^{2} |u|^{2}}
{(1+|u|^{2})^{2}}.
\label{nonsigmamod3}
\end{equation} 

The solutions of the topological solitons can be gained by taking 
the Bogomol'nyi-Prasad-Sommerfield (BPS) bound for the total energy \cite{Bogomolnyi,Prasad}. 
Often, by insisting that the bound is satisfied (called ``saturated"), one can obtain a 
simpler set of partial differential equations to solve, the Bogomol'nyi equations, 
from a square root completion. 
Solutions saturating the bound are called BPS states and their energy is 
proportional to a topological charge that characterizes the solitons. 
Here we summarize the properties of the BPS saturated solutions of the topological solitons. 

\subsection{Vortex}
In the case of $m_{\sigma}=0$ in Eq.~(\ref{nsigmam}), the hamiltonian of the system has O(3) symmetry. 
Since the symmetry of the ground state is broken to O(2), the order parameter space is 
$G/H$ = O(3)/O(2) $\simeq S^2$. Then, the second homotopy group is nontrivial as $\pi_{2}(S^2) = {\bf Z}$. 
This suggests the presence of point-like defects such as monopoles and two-dimensional 
non-singular defects such as ``2D skyrmions" (coreless vortex) \cite{Leslie,Choi}, 
because the former configuration can be mapped to the
latter through the stereographic projection. 

First, we derive the analytic solutions of the coreless vortices by taking the BPS bound. 
We restrict ourselves to consider static solutions which are translationally 
invariant along the $z$-axis. 
The total energy can be written as 
\begin{eqnarray}
E = \int d^2x \frac{\partial_x u \partial_x u^{\ast}  + \partial_y u \partial_y u^{\ast}}{(1+|u|^2)^2} \nonumber \\
= 2 \int d^2x \frac{ | \partial_w u |^2  + | \partial_{\bar{w}} u |^2 }{(1+|u|^2)^2}, 
\label{sigmavorenergy}
\end{eqnarray}
where we have introduced $w=x+iy$, $\partial_w= (\partial_x - i \partial_y) / 2 $ 
and $\partial_{\bar{w}}= (\partial_x + i \partial_y) / 2$. 
The topological charge $T_{\rm v}$ is given by the topological degree of the 
map $u$: $R^2 \to S^2$. By considering the normalized area element of $S^2$, 
the degree of $u$ is given by 
\begin{eqnarray}
T_{\rm v}=\frac{i}{2\pi} \int_{S^2} \frac{du \wedge du^{\ast}}{(1+|u|^2)^2} = 
\frac{1}{\pi} \int d^2 x \frac{|\partial_w u|^2 - |\partial_{\bar{w}} u|^2}{(1+|u|^2)^2} . 
\label{sigmavorcharge}
\end{eqnarray}  
The topological charge of the 2D skyrmion is given by $T_{\rm v} \in {\bf Z} = \pi_2 (S^2)$, giving
the winding number of vortices passing through a certain $z = $ const. plane. 
Inserting Eq.~(\ref{sigmavorcharge}) to Eq.~(\ref{sigmavorenergy}), we find that the total energy 
can be written as the sum of the topological charge and a positive correction:
\begin{eqnarray}
E = 2 \pi T_{\rm v} + 4 \int d^2x  \frac{ | \partial_{\bar{w}} u |^2 }{(1+|u|^2)^2}. 
\end{eqnarray}
Thus the energy is bounded by the topological charge $E \geq 2 \pi T_{\rm v}$ and 
the equality holds only if 
\begin{equation}
\partial_{\bar{w}} u = 0. \label{sigmavorBeq}
\end{equation}
This equation is called a Bogomol'nyi equation. It is a first order equation whose 
solution gives a field configuration with a minimal energy within a fixed topological sector $T_{\rm v}$. 
Equation (\ref{sigmavorBeq}) also shows that $u$ is a holomorphic function of $w$ only. 
Note that $u$ is allowed to have a pole at any point $w=w_i$ because its image on the target $S^2$ 
is just the north or south pole. The requirement that the total energy is finite, together with the boundary 
condition that $u$ has a definite limit as $|w| \to \infty$, forces $u$ to be 
a rational map: 
\begin{equation}
u_{\rm v}(w) = \frac{f(w)}{g(w)} = \frac{\prod_{i=1}^{N_{n}} (w - w^{n}_{i}) }{\prod_{i=1}^{N_{s}} (w - w^{s}_{i})}, 
\label{sigmamultivorsol}
\end{equation}
where $f$ and $g$ are polynomials in $w$ with no common factors. 
This solution gives the vortex configuration, in which $f(w)$ and $g(w)$ 
represent $N_{n}$ vortices (north poles) and $N_{s}$ antivortices (south poles), respectively. 
The positions of the vortices are denoted by $ w^{n}_{i}$ and $ w^{s}_{i}$. 
Note that the total energy does not depend on the form of the solution, 
but only on the topological charges. 
In the NL$\sigma$M, the energy is independent of the vortex positions $ w^{n,s}_{i}$; 
in other words, there works no static interaction between vortices.

\begin{figure}
\begin{center}
\includegraphics[width=0.98 \linewidth,keepaspectratio]{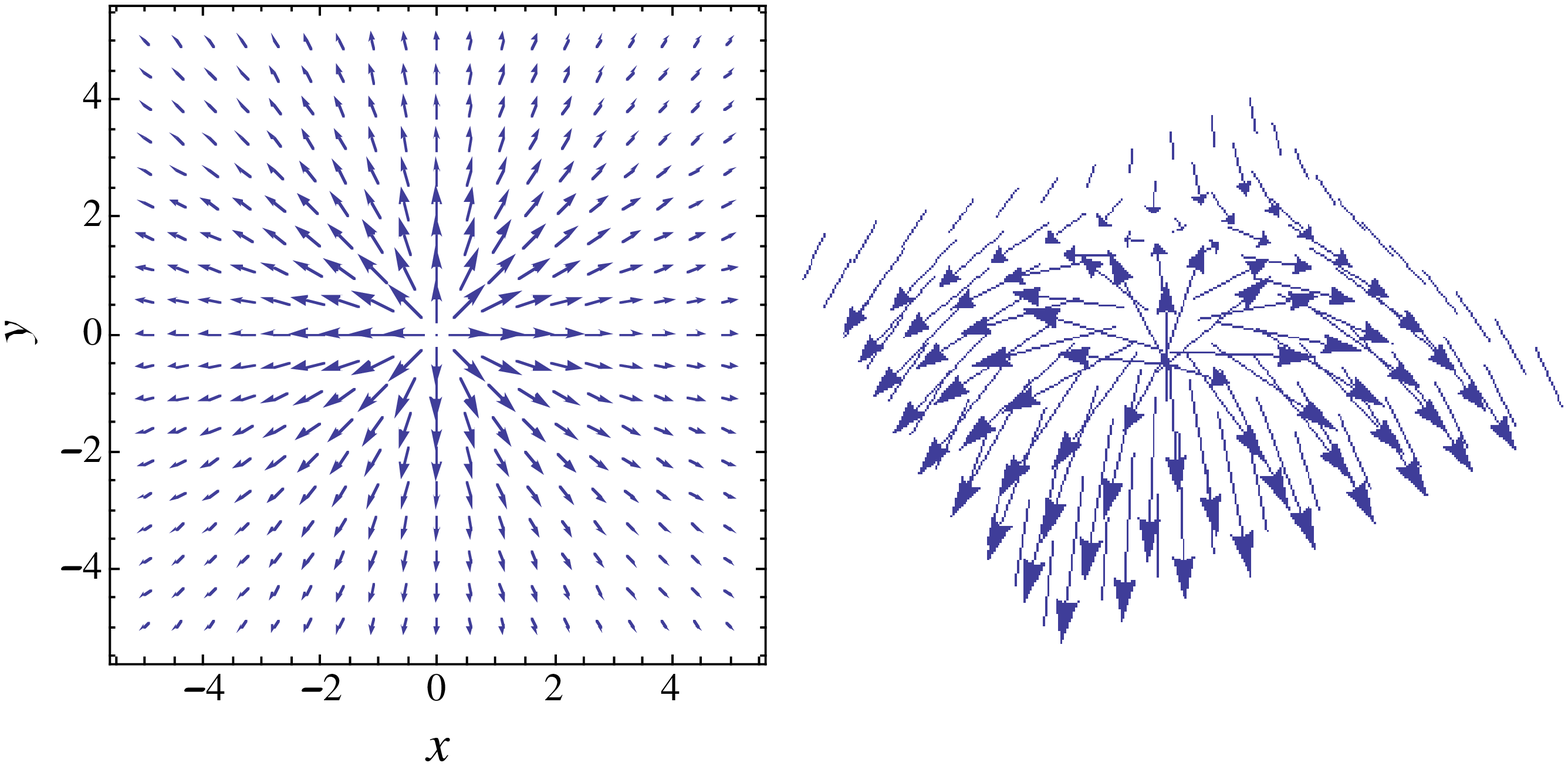}
\end{center}
\caption{(Color online) The spin profile of a coreless vortex. 
The spin component is given by $(s_x,s_y,s_z) = (2x/(1+r^2), 2y/(1+r^2), (1-r^2)/(1+r^2)) $. } 
\label{Figvortex}
\end{figure}
Figure \ref{Figvortex} represents the profile of the spin field for the simple
vortex solution $u_{\rm v} (w) = w$. 
The spin orients upwards at the center ($s_z \rightarrow 1$ as $|w| \rightarrow 0$) 
and it continuously rotates from up to down 
as it moves outward radially ($s_z \rightarrow -1$ as $|w| \rightarrow \infty$). 
The spin configuration of this continuous (coreless) vortex is known as a lump in field theory \cite{Belavin}, 
an Anderson-Toulouse vortex in superfluid $^3$He \cite{AndersonToulouse}, 
or a 2D skyrmion in spinor BECs \cite{Leslie,Choi}. 

\subsection{Domain wall}\label{dimainsigmasec}
Next, let us consider the case for $m_{\sigma} \neq 0$. Assume that the field configuration 
${\bf s}$ interpolates between two degenerate ground state, so that 
$s_{z}=-1$ for $z \to -\infty$ and $s_{z}=1$ for $z \to \infty$. 
Then, we can have the domain wall between the two ground states. 
For such a configuration, we obtain the BPS bound of the energy as 
\begin{eqnarray}
E &=& \int dz \frac{ \partial_z u \partial_z u^{\ast} + m_{\sigma}^2 |u|^2}{(1+|u|^2)^2} \nonumber \\
&=& \int dz \biggl[ \frac{|\partial_z u +  m_{\sigma} u|^2}{(1+|u|^2)^2} 
- \frac{ m_{\sigma} (u^{\ast}\partial_z u + u \partial_z u^{\ast})}{(1+|u|^2)^2} \biggr].
\end{eqnarray}
Here,  the second term corresponds to the charge of the domain wall as
\begin{eqnarray}
T_{\rm w} &=& - m_{\sigma} \int_{-\infty}^{+\infty} dz  \frac{(u^{\ast} \partial_z u + u \partial_z u^{\ast})}{(1+|u|^2)^2} \nonumber \\
&=& \frac{m_{\sigma}}{2} \int_{-\infty}^{+\infty} dz \partial_z s_z 
= \frac{m_{\sigma}}{2} \left[ s_z \right]^{z=+\infty}_{z=-\infty} .
\label{sigmawalltension}
\end{eqnarray}
Under the above boundary condition, the charge becomes $T_{\rm w} = + m_{\sigma}$.
The energy is bounded by the domain wall charge $E \geq T_{\rm w} $ and the saturated solution 
satisfies the Bogomol'nyi equation 
\begin{equation}
\partial_z u + m_{\sigma}  u = 0. 
\label{sigmawallbogoeq}
\end{equation}
Then, we can obtain the BPS wall (kink) solutions
\begin{equation}
u_{\rm w} = e^{- m_{\sigma}  (z-z_{0}) - i \phi_{0}}, 
\label{sigmawallsolutions}
\end{equation}
or, in terms of $s_z$, we have 
\begin{equation}
s_{z} 
= \tanh [m_{\sigma} (z-z_0)].
\label{sigmawallsolutionssz}
\end{equation}
We can also consider the related solution with $T_{\rm w} = - m_{\sigma}$, called 
anti-wall (antikink), and is obtained by making the replacement $u \to u^{-1}$ ($s_{z} \to -s_{z}$). 
Here, $z_0$ represents the position of the flat domain wall ($s_z = 0$) whose 
transverse shift causes the Nambu-Goldstone mode due to 
breaking of the translational invariance. The phase $-\phi_0$ corresponds 
to the azimuthal angle of the pseudospin ${\bf s}$, 
causing the breaking of the global U(1) symmetry {\it locally} along the wall. 
By promoting these two variables to dynamical 
fields as $z_0 \to z_0(x,y,t)$ and $\phi_0 \to \phi_0(x,y,t)$, we can construct an effective theory of the domain wall. 
In the relativistic context, the low-energy dynamics of a single domain in the NL$\sigma$M wall 
can be described by the DBI action \cite{Gauntlett,Shifman}, 
where the {\it local} U(1) gauge fields living on the wall are created by the dual transformation of the 
localized zero mode of the phase $\phi_0$. 
This is the important ground why the domain wall in the NL$\sigma$M can be 
identified as an analog of a D-brane \cite{Gauntlett,Shifman,Sakai}. 

\subsection{Wall-vortex complexes: D-brane solitons}\label{dimvorainsigmasec}
By combining the above two solutions of the topological solitons, 
we can construct the solutions 
in which vortices and domain walls are coexist. 
For a fixed topological sector, namely, for vortices (a domain wall) parallel (perpendicular) 
to the $z$-axis, the total energy Eq.~(\ref{nonsigmamod3}) can be bounded with the 
topological charge as 
\begin{eqnarray}
E \geq |T_{\rm w}| + 2 \pi |T_{\rm v}|.
\end{eqnarray}
The BPS states can be represented with the form of the separated coordinate variables \cite{Isozumi}
\begin{eqnarray}
u({\bf r}) &=& u_{\rm v} (w) u_{\rm w} (z) \nonumber \\
&=& \frac{\prod_{i=1}^{N_{n}} (w - w^{n}_{i}) }{\prod_{i=1}^{N_{s}} (w - w^{s}_{i})} e^{- m_{\sigma}  (z-z_{0}) - i \phi_{0}} ,
\label{wallvortexcompl}
\end{eqnarray}
where $u_{\rm v}$ and $u_{\rm w}$ are satisfied with the 
Bogomol'nyi equations (\ref{sigmavorBeq}) and (\ref{sigmawallbogoeq}), and 
the forms of the solutions are given by Eq.~(\ref{sigmamultivorsol}) and Eq.~(\ref{sigmawallsolutions}). 

\begin{figure}
\begin{center}
\includegraphics[width=0.98 \linewidth,keepaspectratio]{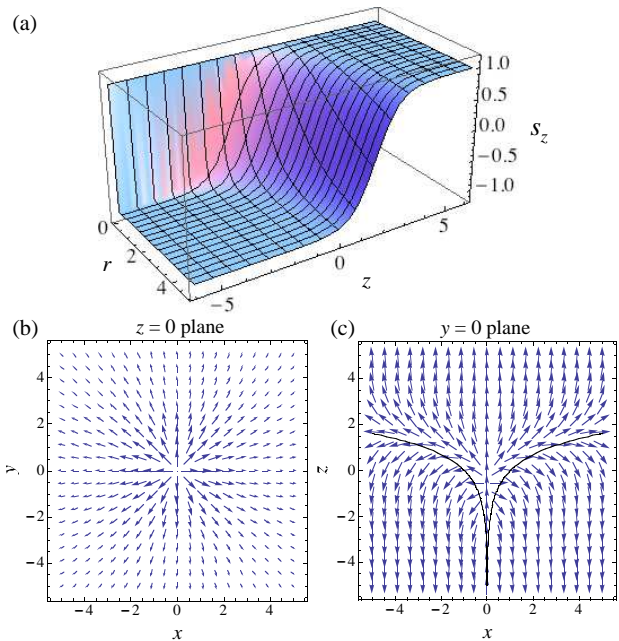}
\end{center}
\caption{(Color online) 
The simplest wall-vortex soliton, where a single vortex along the $z$-axis for $z<0$ 
is connected to the domain wall. The spin profile is explicitly given by 
Eq.~(\ref{spinsimple}) with $m_{\sigma} = 1$. (a) The profile of $s_z$. The spin texture 
of (b) $z=0$ plane and (c) $y=0$ plane. In (c), the wall position given by 
Eq.~(\ref{logbending1}) is shown by solid curve.} 
\label{Fig2sigma}
\end{figure}
To see the properties of the typical solutions, we depict the profile for the solutions 
of a single vortex and a single wall, written as
\begin{equation}
u (w,z) = e^{-m_{\sigma} z} w. 
\label{onewallvortexcomp}
\end{equation}
Here, we choose $z_{0}=0$, $\phi_{0}=0$, 
$N_{n}=1$, $N_{s}=0$, and $w_{1}^{n}=0$. 
This is the simplest composite wall-vortex solution. 
The corresponding spin profile is written as 
\begin{eqnarray}
\left(
\begin{array}{c}
s_{x} \\
 s_{y} \\
  s_{z} 
  \end{array} \right) 
= \left(
\begin{array}{c}
 \dfrac{2 x e^{-m_{\sigma} z}}{1+|w|^2 e^{-2 m_{\sigma} z}} \\ 
 \dfrac{2 y e^{-m_{\sigma} z}}{1+|w|^2 e^{-2 m_{\sigma} z}} \\
\dfrac{1-|w|^2 e^{-2 m_{\sigma} z}}{1+|w|^2 e^{-2 m_{\sigma} z}} 
\end{array}
\right), 
 \label{spinsimple}
\end{eqnarray} 
which is shown in Fig.~\ref{Fig2sigma}. For fixed $|w|=r$ we have a domain wall solution 
along $z$-direction but for fixed $z$ we have a vortex configuration in the $x$-$y$ plain. 
For fixed $z$ the coreless vortex has a scale size $\exp(m_{\sigma} z)$; thus the size 
becomes infinity (zero) as $z \to + \infty$ ($z \to -\infty$). 
The wall position, i.e. the isosurface of $s_z = 0$, is described by a logarithmic function 
\begin{equation}
z=\frac{1}{m_{\sigma}} \ln |w|.
\label{logbending1}
\end{equation}
as seen in Fig.~\ref{Fig2sigma}(c). This situation is equivalent to the logarithmic 
bending when a membrane with a tension $T$ is pulled by a pin at $r=0$, 
where the profile of the membrane is given by $z=( \ln r) / T$ as a problem of mechanics. 
In the BPS solution, the tension of the domain wall is $T_{\rm w} = m_{\sigma}$ 
as shown above. 

\begin{figure}
\begin{center}
\includegraphics[width=0.98 \linewidth,keepaspectratio]{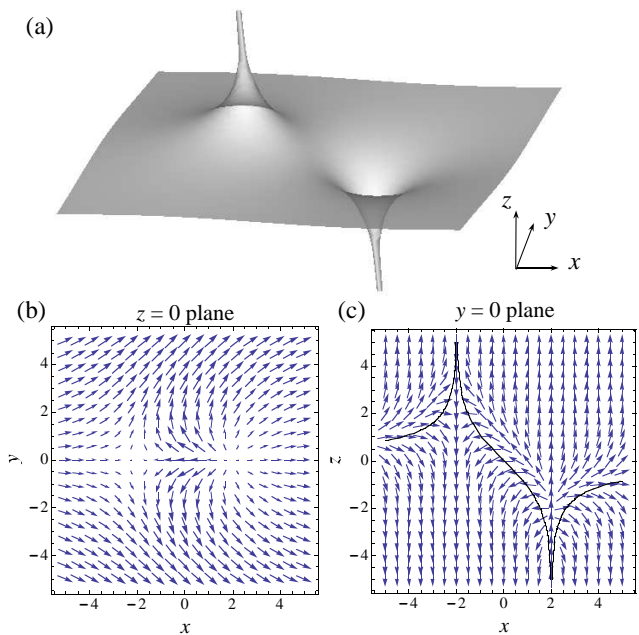}
\end{center}
\caption{(Color online) The D-brane soliton to which two vortices attach.
(a) The isosurface of $s_z = 0$ for the solution Eq.~(\ref{walltwovortexcomp}) of the NL$\sigma$M. 
The corresponding spin textures ${\bf s}$ in the $z = 0$
plane and $y = 0$ plane are shown in (b) and (c), respectively. 
In (c), the wall position given by 
Eq.~(\ref{walltwovortexcompposi}) is shown by solid curve.
The wall becomes asymptotically flat due to the balance between the tensions
of the attached vortices. } 
\label{Fig3sigma}
\end{figure}
We can construct solutions in which an arbitrary number of vortices 
are connected to a domain wall from Eq.~(\ref{wallvortexcompl}), because 
of the absence of the static interaction between vortices. 
Figure \ref{Fig3sigma} shows a solution in 
which two straight vortices along the $z$-axis are connected to the wall on both sides. 
We assume that the vortices have ends at the positions 
$w=x_0$ and $-x_0$ upon the wall. Then, 
 \begin{equation}
u (w, z) = e^{- m_{\sigma} z} \frac{w-x_0}{w+x_0}.
\label{walltwovortexcomp}
\end{equation}
The wall position is given by 
\begin{equation}
z= \frac{1}{m_{\sigma}} \ln \left| \frac{w-x_0}{w+x_0} \right|,
\label{walltwovortexcompposi}
\end{equation}
which becomes asymptotically flat ($z \to 0$) for $|w| \to \infty$. 

It is instructive here to understand the wall configuration Eqs. (\ref{logbending1}) 
and (\ref{walltwovortexcompposi}) in terms of tension of the vortices and walls schematically. 
The wall bending can be interpreted to be caused by the tension of vortex attached to the wall. 
For the case of Eq. (\ref{walltwovortexcompposi}), the tensional forces by the two vortices balance 
and the equilibrium position of the wall is well-defined as $z=0$ 
asymptotically. However, for Eq. (\ref{logbending1}) 
the position is ill-defined since the force is unbalanced in the presence of a vortex only 
in the one side of the wall. Similarly, the wall has an equilibrium position $z=z_0$ when 
the number of vortices in one side equals to that in the other side, namely $N_n=N_s$ 
in Eq. (\ref{wallvortexcompl}). 
The stiffness of the wall is represented by the coefficient $m_{\sigma}^{-1}$ in Eqs. (\ref{logbending1}) 
and (\ref{walltwovortexcompposi}) with the wall tension $T_{\rm w} = m_{\sigma}$. 
Therefore, the wall is more flexible as the wall tension $m_{\sigma}$ decreases.
 
It should be mentioned here why this wall-vortex composite soliton has been 
referred to as ``D-brane" soliton in the relativistic theory \cite{Gauntlett,Shifman}. 
On a single D-brane, the Abelian gauge theory is realized. 
The present domain wall has a localized U(1) Nambu-Goldstone mode and it can be
rewritten as the U(1) gauge field on the wall, which is a necessary degree of freedom for the 
Dirac-Born-Infeld (DBI) action of a D-brane. Gauntlett {\it et al}. \cite{Gauntlett} have shown 
that Eq.~(\ref{wallvortexcompl}) reproduces the ``BIon" solutions of the DBI action for
D-branes in string theory by constructing an effective theory of the domain-wall world
volume with collective coordinates $z_0(x,y,t)$ and $\phi_0(x,y,t)$ in $u_w (z)$, 
where $\phi_0$ is periodically identified as $\phi_{0} \rightarrow \phi_{0}+2\pi$. 
In the relativistic theory,  the low energy effective action for these collective coordinates 
is given by 
\begin{equation}
I = - T_{\rm w} \int d^3 \xi \sqrt{-{\rm det} (G_{ij} + \partial_i \phi_{0} \partial_j \phi_{0})}, 
\end{equation}
where $(\xi^0, \xi^1, \xi^2) = (t, x, y)$, and $G_{ij} = \eta_{ij} + \partial_{i} z_0 \partial_j z_0$ with 
$\eta_{ij} = {\rm diag}(-1,1,1)$ is the metric induced from the Minkowskii metric for 
a deformed membrane.
Using the localized phase $\phi_0$, we can introduce the $U(1)$ gauge field $A_{j}$ by taking a dual as 
\begin{equation}
\partial_{i} \phi_{0} = \epsilon_{ijk} \partial_{j} A_{k}. 
\end{equation}
The effective action of $z_{0}$ and $A_{i}$ corresponds to the so-called 
DBI action of the D2-brane: 
\begin{equation}
I = - T_{\rm w} \int d^3 \xi \sqrt{-{\rm det} (G_{ij} + F_{ij})}, 
\end{equation} 
where $F_{ij} = \partial_i A_j - \partial_j A_i$ is the electromagnetic field strength. 
The solution of this effective theory in the background of a point source 
with an electric charge and a scalar charge 
is known as BIon and its profile is precisely coincident with that of the wall-vortex soliton 
in the NL$\sigma$M \cite{Gauntlett}.
Thus, the endpoints of the vortex lines in the NL$\sigma$M can be seen as electrically 
charged particles within this effective theory \cite{Gibbons:1997xz,Callan}, and 
the domain wall can be seen as a D-brane on which fundamental strings terminate. 
However, the correspondence should be modified in our non-relativistic theory, 
which should be considered in more detail but is beyond the scope of this paper.   

\section{Wall-vortex composite soliton in two-component BECs} \label{numeri}
The mapping into the NL$\sigma$M can allow one to identify the domain wall 
of the two-component BECs as a non-relativistic counterpart of the D-brane soliton. 
Based on the analytic solutions of topological solitons in the simplified NL$\sigma$M, 
we next consider the structure of the wall-vortex composite soliton in trapped two-component 
BECs. The generalized NL$\sigma$M Eq.~(\ref{nonsigmamodBEC}) has additional terms which are absent 
in the original NL$\sigma$M. Here, we discuss the modification of the soliton structure 
in the two-component BEC from the analytical solutions of Eq.~(\ref{wallvortexcompl}). 

For simplicity, we assume the symmetric parameters $m_{1} = m_{2} \equiv m$ 
and $g_{1} = g_{2} \equiv g = 4 \pi \hbar^2 a / m$.
We introduce the external trapping potential of Eq.~(\ref{trap}). 
To nucleate and stabilize the vortices in trapped condensates, 
the system is supposed to be rotated at a rotation frequency ${\bf \Omega} = \Omega {\bf \hat{z}}$. 
In order to compare the numerical results with the previous analytical results directly, 
we introduce the length scale $\xi = \hbar/\sqrt{2mg\rho(0)}$ and the energy scale $\mu = g \rho(0)$, 
where $\rho(0)$ is the total density at the center of the trapping potential and can be estimated 
easily by applying the Thomas-Fermi approximation \cite{Pethickbook}. 
The coupled GP equation derived from the energy functional of Eq.~(\ref{energyfunctio2}) 
in the rotating frame of ${\bf \Omega}$ can be written as 
\begin{eqnarray}
\left[ - \nabla^2 + \tilde{V} + |\Psi_j|^2  + \gamma |\Psi_{k}|^2 -\alpha \tilde{\Omega} L_z \right] \Psi_j= \Psi_j ,
\label{dimles2GPeq}
\end{eqnarray}
where $\alpha = (4 \pi \rho(0) a a_{\rm ho}^2)^{-1}$ with $a_{\rm ho} \equiv \sqrt{\hbar/ m \omega}$ and 
$\gamma \equiv g_{12}/g$.
The wave function has been scaled as $\Psi \to \sqrt{\rho(0)} \Psi$. 
In the following, we confine ourselves to the parameter range $\gamma > 1$ 
for which the phase separation occurs. 
The trapping potential $\tilde{V}$ can be written as 
\begin{eqnarray}
\tilde{V} = \frac{\alpha}{2} \left( \frac{\xi}{a_{\rm ho}} \right)^2 (r^2 + \lambda^2 z^2),
\end{eqnarray}
and the rotation frequency is $\tilde{\Omega} = \Omega/\omega_{\perp}$
The wave functions are normalized as $\int d {\bf r} |\Psi_j|^2 = N_j/ [\xi^3 \rho(0)] $. 
The numerical solutions shown below are calculated by the imaginary time 
propagation of Eq.~(\ref{dimles2GPeq}). 

Note that, in order to realize the configuration as shown in Fig.~\ref{Fig1ponch}, the resulting
domain wall should be perpendicular to the rotation axis. 
Then, it is desirable that the global shape of the condensate is elongated along the rotation 
($z$) axis, because such a configuration minimizes the interface area between the two domains 
to decrease the energy cost due to the surface tension.  
We thus prepare a cigar-shaped trap with $\lambda =1/4$ to reduce the 
interface area and to keep the interface parallel in the $x$-$y$ plane. 
We fix the intra-species s-wave scattering length as $a=5.61$ nm and consider that 
the inter-species one $a_{12}$ is a free parameter in the following. 
The use of the interspecies Feshbach resonance will be crucial for 
realizing such a situation experimentally \cite{Papp}. 


\subsection{Domain wall} 
We first discuss the structure of a domain wall in two-component BECs 
on the basis of the results of the NL$\sigma$M. The domain wall structures have 
been also studied by several authors \cite{Coen,Barankov}. 
The position of the domain wall is defined as a plane on which two components 
have the same amplitude $|\Psi_1| = |\Psi_2|$ ($s_z = 0$). 
Because the trapping potential does not play an essential role 
in the domain wall structure, we consider the homogeneous 
system with $\tilde{V} = 0$ as well as $\tilde{\Omega} = 0$. 
Then the system is characterized by only one parameter $\gamma = g_{12}/g$.

\begin{figure}
\begin{center}
\includegraphics[width=0.98 \linewidth,keepaspectratio]{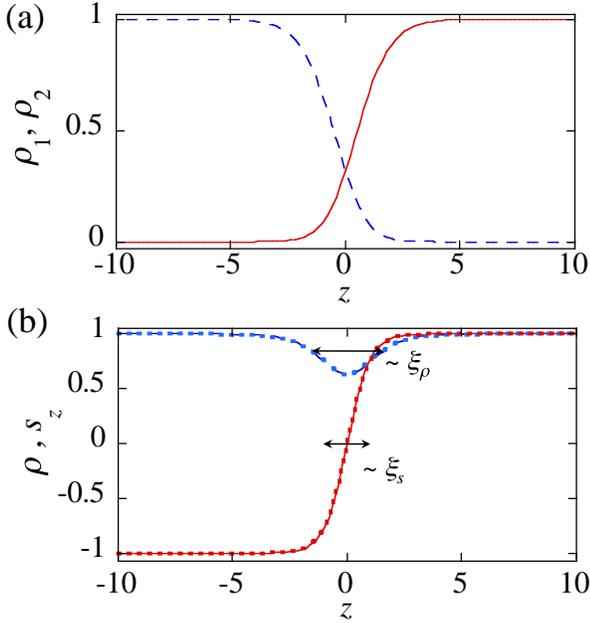}
\end{center}
\caption{(Color online) The numerical solutions of a domain wall in two-component 
BECs for $\gamma=2$ with the boundary conditions 
$(\Psi_1, \Psi_2)^{T} \to (1,0)^{T}$ at $z \to + \infty$, 
and $(\Psi_1, \Psi_2)^{T} \to (0,1)^{T}$ at $z \to - \infty$. 
(a) The density profile of $\rho_1$ (solid curve) and 
$\rho_2$ (dashed curve). (b) The total density $\rho = \rho_1+ \rho_2$ 
(dashed curve) and the $z$-component of the pseudospin $s_z$ (solid curve). 
There are two characteristic length scales $\xi_p$ and $\xi_s$ in the solution. 
The results of the variational ansatz in Eqs.~(\ref{spinzprofile}) 
and (\ref{denansatz}) for the profile of $s_z$ and $\rho$ are also 
plotted by dotted curves, where $\Delta_{\rho}$ in Eq.~(\ref{denansatz}) is a variational parameter.} 
\label{becdomainwall}
\end{figure}
Let us assume that the wall lies in the $z=0$ plane and impose the 
following boundary conditions $(\Psi_1, \Psi_2)^{T} \to (1,0)^{T}$ at $z \to + \infty$, 
and $(\Psi_1, \Psi_2)^{T} \to (0,1)^{T}$ at $z \to - \infty$. 
The typical profile of the domain wall solution is shown in Fig.~\ref{becdomainwall}.
As $z$ increases from negative to positive, the amplitude of $\Psi_2$ decreases 
as it approaches to the domain wall, while 
that of $\Psi_1$ increases from zero with being apart from the domain wall. 
By representing the solution with the total density $\rho = \rho_1 + \rho_2$ and 
the density difference, i.e. the $z$-component of the pseudospin $s_z = (\rho_1 - \rho_2)/\rho$, 
it can be clarified that the domain wall has a two-component structure, as shown in Fig.~\ref{becdomainwall}(b). 
The two length scales can be derived from the generalized NL$\sigma$M Eq.~(\ref{becsimplenlsm}), 
its dimensionless form being given by 
\begin{eqnarray}
\tilde{E} = \int d {\bf r} \biggl[ \left( \nabla \sqrt{\tilde{\rho}} \right)^2 
+ \frac{\tilde{\rho}}{4} \sum_{\alpha} \left( \nabla s_{\alpha} \right)^2 
+ \tilde{\rho} \tilde{v}_{\rm eff}^2 \nonumber \\
+ \frac{1}{2} (\tilde{\rho} -1)^2 
+ \frac{1}{4} m_{\sigma}^2 \tilde{\rho}^2 (1-s_z^2) \biggr].
\label{dimlessGNLSM}
\end{eqnarray}
Here, we put $\tilde{\rho}=\rho/\rho(0)$ and $\tilde{\bf v}_{\rm eff} = (m\xi /\hbar){\bf v}_{\rm eff}$; 
the tilde is omitted in the following discussion. 
By assuming that the system is uniform in $x$- and $y$-directions 
and $v_{\rm eff} = 0$, we consider the system spatially-dependent 
only on the $z$-direction. Using the identity 
\begin{equation}
\sum_{\alpha} (\nabla s_{\alpha})^2 = \frac{1}{1-s_z^2} 
[(\nabla s_z)^2 + (s_y \nabla s_x - s_x \nabla s_y)^2], 
\end{equation}
we can write the energy of the problem as 
\begin{eqnarray}
E_z = \int dz \biggl[ (\partial_z \sqrt{\rho})^2 + \frac{\rho}{4} \frac{(\partial_z s_z)^2}{1-s_z^2} 
+ \frac{1}{2} (\rho-1)^2 \nonumber \\ 
\frac{1}{4} m_{\sigma}^2 \rho^2 (1-s_z^2) \biggr]. 
\label{energy1Dda}
\end{eqnarray}
The stationary solutions of the system satisfy the equations 
\begin{eqnarray}
-\frac{\partial_z^2 \sqrt{\rho}}{\sqrt{\rho}} + \frac{1}{4} \frac{(\partial_z s_z)^2}{1-s_z^2} 
+\frac{1}{2} m_{\sigma}^2 \rho (1-s_z^2) + \rho = 1 ,  \\
\frac{\rho s_z (\partial_z s_z)^2}{(1-s_z^2)^2} + \frac{(\partial_z \rho)(\partial_z s_z) 
+ \rho (\partial_z^2 s_z)}{1-s_z^2} + m_{\sigma}^2 \rho^2 s_z = 0.
\end{eqnarray}
The asymptotic form of the profile can be obtained by linearizing with respect to $\rho$ and $s_z$ 
around the ground state value $\rho=1$ and $s_z = \pm 1$ as 
$\rho \sim 1 - e^{\pm z/\xi_{\rho}} = 1 - e^{\pm \sqrt{2} z}$ 
and $s_z \sim \mp 1 \pm e^{\pm z/\xi_{s}} = \mp 1 \pm e^{\pm \sqrt{2} m_{\sigma} z}$ for $z \to \mp \infty$. 
This gives the characteristic length scales 
$\xi_{\rho} = 1/\sqrt{2}$ and $\xi_s = 1/ (\sqrt{2} m_{\sigma})$ in unit of $\xi$. 
Similar two-component structure can be seen also in the vortex solutions of 
two-component BEC \cite{Eto2BEC}. 
In the strongly segregating limit $g_{12} \to \infty$, the domain wall is characterized by a 
single length scale $\xi_{\rho}$ because $\xi_s$ vanishes. 

Generally, the domain wall solution in the NL$\sigma$M is written as Eq.~(\ref{sigmawallsolutions}). 
In terms of the condensate wave function, the domain wall solution can be written as 
$\Psi_1=f_{1d} (z-z_0) e^{-i\phi_0/2}$ 
and $\Psi_2=f_{2d} (z-z_0) e^{i\phi_0/2}$, where $f_{jd}$ is a real function with the wall center $z_0$ 
and the phase $\phi_0$, which can be identified as the relative phase between two components 
$\phi_0=\phi = \theta_2 -\theta_1$. 
The fixing of $z_0$ is due to the breaking of translational invariance 
by the given wall solution, while $\phi_0$ is due to the breaking of global U(1) 
{\it around the domain wall}, i.e., a narrow overlapping region of the two-component wave functions, 
and consequently there appears a U(1) Nambu-Goldstone mode localized around the wall. 
This feature satisfies a part of the requirement discussed in Sec.\ref{dimainsigmasec} 
and \ref{dimvorainsigmasec} that 
the domain wall in two-component BECs can be referred to as a D-brane 
soliton \cite{KasamatsuD}.

\begin{figure}
\begin{center}
\includegraphics[width=0.98 \linewidth,keepaspectratio]{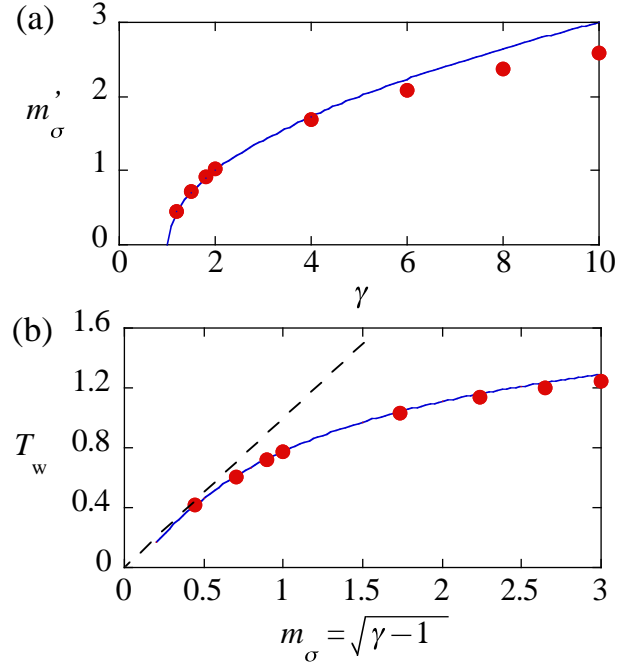}
\end{center}
\caption{(Color online) (a) The relation between $m_{\sigma}'$ and $\gamma$. 
The circles are obtained by the fitting 
of the numerical solution and Eq.~(\ref{spinzprofile}) with the fitting parameter $m_{\sigma}'$. 
We show the result for the $\sigma$-model limit 
$m_{\sigma} = \sqrt{|1-\gamma|}$ by the solid curve for comparison. 
(b) Tension of a domain wall as a function of $m_{\sigma} = \sqrt{|1-\gamma|}$. 
The solid curve represents 
the results calculated from the ansatz of Eqs (\ref{sigmawallsolutionssz}) and (\ref{denansatz}), 
where the energy is optimized with respect to the variational parameter $\Delta_{\rho}$. 
The dots show the results calculated with numerical solutions. The dashed line 
represents the tension $T_{\rm w} = m_{\sigma}$ in the $\sigma$-model limit } 
\label{becdomainwall2}
\end{figure}
It is instructive to consider the analytical form of the domain wall solution. 
The domain wall in the NL$\sigma$M of Eq.~(\ref{nonsigmamod2}) 
under the $g \to \infty$ ($\xi_{\rho} \to 0 $) limit has a single characteristic length $\xi_s$, where 
the profile is given by Eq.~(\ref{sigmawallsolutions}) or (\ref{sigmawallsolutionssz}). 
When the spatial gradient of $\rho$ is small enough for $\gamma \gtrsim 1$,
the profile of $s_z$ in the generalized NL$\sigma$M must follow that in the NL$\sigma$M. 
We find that the domain wall solutions in Fig.~\ref{becdomainwall} also follow 
correctly this profile function with slightly modified mass $m_{\sigma}'$:
\begin{equation}
s_z 
= \tanh (m_{\sigma}' z). 
\label{spinzprofile}
\end{equation}
We make a fit of the numerical solution to Eq.~(\ref{spinzprofile}) 
to extract the fitting value of $m_{\sigma}'$, which is plotted as a function 
of $\gamma$ in Fig.~\ref{becdomainwall2}(a). 
The mass parameter is almost in agreement with the values given in Eq.~(\ref{2becnlsmmass}) 
in the sigma model limit, although it is slightly deviated as $\gamma$ increases. 
Thus, the domain wall in two-component BECs can be regarded as the same solitonic 
object in the original NL$\sigma$M, including the quantitative details of the structure. 

The remaining total density can be described by the ansatz 
\begin{equation}
\rho = 1 - \Delta_{\rho} {\rm sech} \left( \frac{z}{\sqrt{2}\xi_{\rho}} \right) ,
\label{denansatz}
\end{equation}
where $\Delta_{\rho}$ is a variational parameter. According to 
Fig. \ref{becdomainwall2}(a), it is reasonable to put $m_{\sigma}' \approx m_{\sigma}$ 
in Eq. (\ref{spinzprofile}) as $s_z = \tanh (m_{\sigma} z) 
= \tanh (z/\sqrt{2} \xi_s) $; in other words, $s_z$ is assumed to be 
given by Eq. (\ref{sigmawallsolutionssz}) with $z_0 = 0$.
Inserting Eqs.~(\ref{sigmawallsolutionssz}) and (\ref{denansatz}) into the energy 
of Eq. (\ref{energy1Dda}) and minimizing the energy with respect 
to $\Delta_{\rho}$, we obtain the domain wall profile semi-analytically as shown in 
Fig.~\ref{becdomainwall}(b). The ansatz of Eqs.~(\ref{sigmawallsolutionssz}) 
and (\ref{denansatz}) agrees with the numerical result almost perfectly.  
The optimized energy corresponds to the tension of a domain wall $E_{z}^{\rm min} = T_{\rm w}$, 
which is an extended version of Eq. (\ref{sigmawalltension}) for the two-component BECs. 
The tension is simply given by $T_{\rm w} = m_{\sigma}$ in the sigma model case, while it is significantly 
reduced in the BEC case because of the additional contribution of the total density, as shown in 
Fig.~\ref{becdomainwall2}(b).

\subsection{Axisymmetric structure of wall-vortex complex} 
We next consider the axisymmetric wall-vortex soliton in trapped two-component BECs. 
As shown in Fig.~\ref{Fig1ponch}, the $\Psi_1$ ($\Psi_2$) domain is placed 
at $z>0$ $(z<0)$ and each component is assumed to have a straight vortex line at the center. 
The axisymmetric solution $\Psi_j = f_j(r,z)e^{i n_j \theta}$ with the real function 
$f_j$, the polar angle $\theta$ and the vortex winding number $n_j$ can 
be obtained by numerically solving the coupled GP equations: 
\begin{eqnarray}
\left[ - \left( \frac{\partial^2}{\partial r^2} + \frac{1}{r} \frac{\partial}{\partial r} +\frac{\partial^2 }{\partial z^2}  - \frac{n_{j}}{r^2} \right) 
+ \tilde{V} + f_j^2 + \gamma f_{k}^2 \right]  f_j \nonumber \\ 
= f_j.
\label{axisymtwoGP}
\end{eqnarray}
The parameters are the ratio of the coupling constants $\gamma \equiv g_{12}/g$ 
and the winding number $n_{j}$. 
Here, we assume for simplicity that both components have the same particle number $N$, 
setting $m=m_{\rm ^{87}Rb}$, $\omega=20\times 2 \pi$ Hz, and $N=10^{5}$. 

Before proceeding the discussion, we give some notes on the numerical solutions. 
In the energy-minimization process of the numerical simulations, the chemical potential is 
usually fixed in a homogeneous problem without a trapping potential, 
so that the particle number of each component is not conserved. 
Then, the pressure difference between two components, originated from  
the asymmetry of the solution, leads to the decrease in the population of the 
energetically unfavorable vortical component during the imaginary time evolution. 
Eventually, the vortex-free component fills all space as a final equilibrium solution. 
To obtain the desired solution, we have to adjust the chemical potential difference to 
balance the pressure between the two components, which is a troublesome task. 
Thus we make the numerical minimization by fixing the particle number in each 
component in the presence of the trapping potential, which is an experimentally 
relevant situation. 

\begin{figure}
\begin{center}
\includegraphics[width=0.98 \linewidth,keepaspectratio]{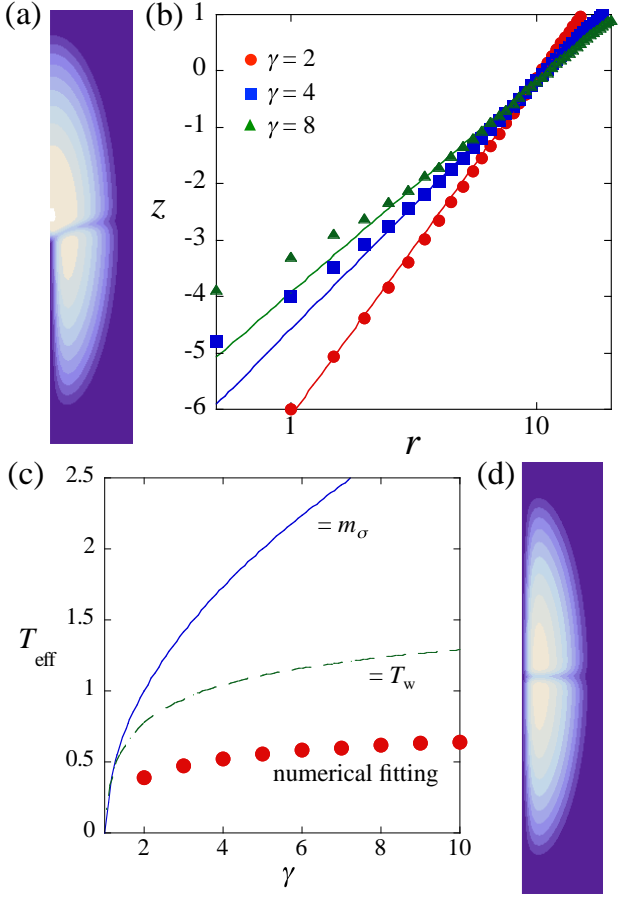}
\end{center}
\caption{(Color online) The numerical solutions of the axisymmetric wall-vortex soliton obtained 
by the GP equation. (a) The profile of the total density for $(n_1,n_2)=(0,1)$ and $\gamma=2$ 
(b) Semi-log plot of the wall position for several values of $\gamma=g_{12}/g$. Each plot 
can be fitted by the logarithmic function as $z=A \ln r + B$ with $A=2.59$, 1.91, and 1.62 
for $\gamma=2$, 4, and 8, respectively, where the fitting is made by using the data 
for $r>5$ to avoid the contribution of the singular vortex core. 
(c) The effective tension determined from the relation $T_{\rm eff} = A^{-1}$ (see Eq. (\ref{becwallpositry}))
as a function of $\gamma$. The circles are obtained from the numerical fitting.
The solid curve corresponds to the $\sigma$-model limit: 
$T_{\rm eff} = m_{\sigma} = \sqrt{|1-\gamma|}$, while the dashed curve 
is domain wall tension obtained by the variational approach, 
shown in the solid curve in Fig. \ref{becdomainwall2}(b). 
(d) The profile of the total density for $(n_1,n_2)=(1,1)$.} 
\label{Sigmafitting}
\end{figure}
Figure \ref{Sigmafitting} (a) and (b) show the profile of the total density $\rho=\rho_1 + \rho_2$ 
for $\gamma=2$ and the positions of domain wall, 
namely $s_z = 0$ ($|\Psi_1| = |\Psi_2|$) for several values of $\gamma$, respectively. 
The vortex in the $\Psi_{2}$-component near the domain wall forms a coreless vortex, 
where its core is filled by the density of the $\Psi_{1}$-component and transforms 
into a singular vortex with distance from the domain wall. 
Thus, the total density $\rho$ is reduced at the position of the domain wall 
and vanishes at the singular vortex core around $r=0$ for $z<-z_0$. 
The appearance of the singular core can be understood from the generalized 
NL$\sigma$M. When we give the phases $\theta_1 = 0$ and $\theta_2 = \theta$ 
for the case of Fig.~\ref{Sigmafitting}(a), for example, we obtain 
\begin{equation}
\tilde{v}_{\rm eff}^2 = \frac{1}{4} (\nabla \theta_2)^2 (1-s_z)^2 = \frac{1}{4r^2} (1-s_z)^2. 
\end{equation}
This kinetic energy density vanishes in the $\Psi_1$ ($s_z = 1$) domain, while 
it contributes to the energy as $\tilde{\rho} r^{-2}$ in the $\Psi_2$ ($s_z = -1$) domain. 
The latter divergent contribution makes the singular vortex core in the 
$\Psi_2$-component around $r=0$. 

This inhomogeneity of $\rho$ implies that the assumption of the uniform total 
density to derive the NL$\sigma$M in Eq.~(\ref{nonsigmamod2}) is not good 
and the solution is expected to be deviated from Eq.~(\ref{onewallvortexcomp}). 
Nevertheless, the spin texture of this solution is almost identical 
to that in Fig.~\ref{Fig2sigma}. 
The plot of the wall position in Fig.~\ref{Sigmafitting} (b) can be well fitted by 
the logarithmic function $z=A \ln r + B$, as expected from Eq.~(\ref{logbending1}). 
Thus, the qualitative structural feature of the wall-vortex composite soliton 
in two-component BECs is not changed from the BPS solutions of the NL$\sigma$M. 
According to Eq.~(\ref{logbending1}), we apply an analogy of a pulled membrane 
to this situation and extract the effective tension $T_{\rm eff}$ 
of the domain wall from the numerical fitting with the relation
\begin{equation}
z-B = \frac{1}{T_{\rm eff}} \ln |w|.
\label{becwallpositry}
\end{equation} 
Here, the influence of the vortex associated with ${\bf v}_{\rm eff}$ is neglected. 
As shown in Fig.~\ref{Sigmafitting}(c), the value of the effective 
tension $T_{\rm eff} = A^{-1}$ is significantly reduced from not only 
$T_{\rm w} = m_{\sigma} = \sqrt{|1-\gamma|}$ of Eq.~(\ref{nonsigmamod2}) 
in the $\sigma$-model limit but also $T_{\rm w}$ of the BEC domain wall 
in Fig.~\ref{becdomainwall2}(b). 
This means that the domain wall in this composite soliton can 
be more flexible than that in a single domain wall. 
This further reduction of the tension may be attributable to the following effects: 
(i) The rotational flow of a vortex causes the density inhomogeneity, 
where the density changes as $\rho_2 \sim r^2/(2+r^2)$ around the singular vortex core. 
The density difference between $\rho_1$ and $\rho_2$ ($\rho_1 > \rho_2$ near the wall) 
can enhance pressure from $\Psi_1$ to $\Psi_2$ and lead to bend the wall more flexibly. 
(ii) The trapping potential also gives rise to the density inhomogeneity 
and the pressure balance can be modified radially. The latter is 
probably minor effect because Fig.~\ref{Sigmafitting}(b) shows that 
the wall is well-fitted logarithmically even for large $r$. 
Also, the effective tension is almost equal to that of the solution calculated in 
the uniform system, although $\mu_1 \neq \mu_2$ \cite{Kasamatsuissue}. 

Figure \ref{Sigmafitting}(d) represents the profile of the total density for $(n_1,n_2) = (1,1)$. 
Because of the balance of the vortex tension, the domain wall becomes flat. 
This situation corresponds to $x_0 = 0$ in Eq.~(\ref{walltwovortexcomp}) in the sigma model. 
There are two singular vortices which has infinitely thin distribution 
($\delta$-functional form) of the vorticity; thus we have only the domain wall structure 
because the relative phase between the two components is uniform everywhere. 

\begin{figure}
\begin{center}
\includegraphics[width=0.98 \linewidth,keepaspectratio]{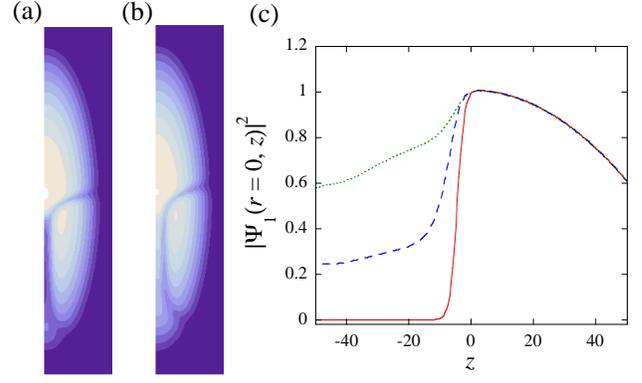}
\end{center}
\caption{(Color online) The numerical solutions of the axisymmetric wall-vortex soliton obtained 
by the GP equation for $\gamma=2$. The profile of the total density for (a) $(n_1,n_2)=(0,2)$ 
and (b) $(n_1,n_2)=(0,3)$ (c) The profile of $\rho_1$ at $r=0$ as a function 
of $z$ for $n_2=1$ (solid curve), 2 (dashed curve), and 3 (dotted curve).} 
\label{Sigmafitting2}
\end{figure}
Figure \ref{Sigmafitting2} represents the solution for $(n_1, n_2) = (0,2)$ 
and $(n_1, n_2) = (0,3)$. In these cases, the size of the vortex core extends radially, 
and the core is filled by the $\Psi_1$-component to be identified as the 
coreless vortex. According to the BPS solution of the NL$\sigma$M 
$u(w,z) =  e^{-m_{\sigma} z} w^{n_2}$, the position of the domain wall 
is expected to become 
\begin{equation}
z -B \simeq \frac{n_2}{T_{\rm eff}} \ln |w|. 
\end{equation}
The logarithmic fitting $z=A \ln r + B$ (for $r > 5$) of these solutions 
shows $A=$2.59, 4.44, and 6.20 for $n_2=$1, 2, and 3, respectively. 
This is fairly agreement with the property of the NL$\sigma$M solution 
but its increase is lower than the expected linear dependance. 
This means that the tensile force for a domain wall pulled by two vortices
is weaker than that of two BPS vortices. 

Note that in the case with $n_2>1$ in Fig.~\ref{Sigmafitting2}, the total density does not vanish 
at the vortex core, as seen in Fig.~\ref{Sigmafitting2} (c). Since the core size of the 
multiply quantized vortex becomes large with increasing $n_2$ like $\sim [r^2/(2+r^2)]^{n_2}$, 
the density of the vortex-free component can enter the core easily. 
On the other hand, the vortex core for $n_2 = 1$ is apparently singular without density. 
This indicates that there is a critical core size that allows the filling of the density inside the core. 
From different points of view, there is a critical ratio of the chemical potential $(\mu_2/\mu_1)_c$ 
that determines whether the vortex core can be filled with the other non-vortex component 
for a given $\gamma$ \cite{Takeuchiprelim}. The 2D simulation shows that 
the optimized $(n_1,n_2) = (0,1)$ 
vortex state for $\mu_1 = \mu_2 =1$ and $\gamma=2$ is actually characterized by the empty 
vortex core. 

\subsection{Non-axisymmetric structure: a wall with multiple vortices} 
Next, we remove the axisymmetric condition and calculate the equilibrium state by the 
imaginary time propagation of Eq.~(\ref{dimles2GPeq}) in full 3D space 
from a suitably prepared initial configurations. To realize the final equilibrium 
configuration as shown in Fig.~\ref{Fig1ponch}, we prepare the phase separated 
state in which $\Psi_1$ ($\Psi_2$)-domains with some phase singularities (seeds of vortices) 
are located in the $z>0$ $(<0)$ region as the initial state of the calculation. 

\begin{figure*}
\begin{center}
\includegraphics[width=0.98 \linewidth,keepaspectratio]{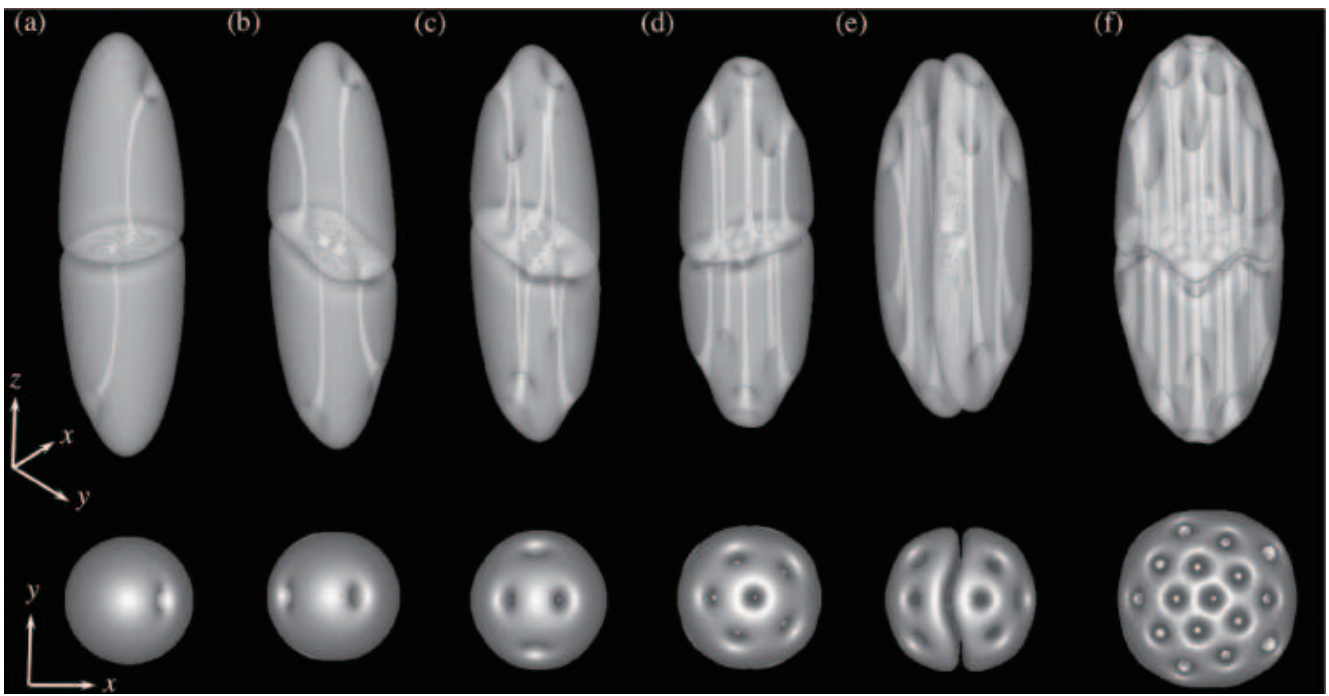}
\end{center}
\caption{The configuration of a wall connecting multiple vortices in trapped BECs. 
The panels show the profile of the density difference $|\rho_1-\rho_2| \propto |s_z|$ 
(isosurface of $|\rho_1-\rho_2|=0.0005$) 
for $\tilde{\Omega}=$ (a) 0.4, (b) 0.5, (c) 0.6, (d) 0.7, and (e) 0.8. 
The parameter values used are $\lambda = 1/4$, $N = 5.5 \times 10^4$, and $a_{12} = 2 a$. 
The lower panels show the top view of the upper panels. In (f), we show the equilibrium 
structure for $\tilde{\Omega} = 0.8$, $\lambda = 1/4$, and $a_{12} = 2 a$ but the particle 
number $N=1.65 \times 10^5$ which is as three times as that in (e). } 
\label{numericalsolu}
\end{figure*}

The panels of Fig.~\ref{numericalsolu} show the 3D distributions of the density difference 
$|\rho_{1}-\rho_{2}| \propto |s_z|$ of the equilibrium state for several $\Omega$; 
this presentation is suitable to visualize the region of the vortex core 
and the domain wall (surface of $\rho_{1} = \rho_{2}$). 
This configuration is energetically stable since it is obtained 
by imaginary time propagation. For $\tilde{\Omega}=0.40$ we obtain the $(n_1,n_2) = (1,1)$ state. 
Contrary to the axisymmetric structure of Fig.~\ref{Sigmafitting}(c), the end point of the vortices 
in each component is spontaneously displaced 
from the center, corresponding to $x_{0} \neq 0$ in Eq.~(\ref{walltwovortexcomp}). 
While the energy is independent of $x_{0}$ in the BPS solution of the NL$\sigma$M, 
this displacement is due to the fact that the vorticity should be distributed 
broadly near the domain wall so as to reduce the associated kinetic energy 
as well as to reduce the gradient of $\rho$. 
Also, the vortex line is slightly bent due to the elongated trapping potential \cite{GarciaRipoll}. 
Because our calculation uses the same rotation frequency $\tilde{\Omega}$ for both components, 
the number of the nucleating vortices should be the same for the both components \cite{tyuu}. 

When the rotation is further increased, multiple vortices form a lattice 
in each component. Then, the domain wall begins to incline from the $z=0$ 
plane [Fig.~\ref{numericalsolu}(b)-(d)] and eventually becomes parallel to the 
rotation axis [Fig.~\ref{numericalsolu}(e)], even though the interface area (energy) increases. 
This is a vortex sheet structure \cite{Kasamatsu3}. 
The reason why this vortex sheet structure is preferred is due to the 
fact that the absorption of the vortices into the domain wall leads to 
the decrease in gradient energy of the singular vortex cores. 
This effect is also absent in the composite solitons of the NL$\sigma$M, 
which is free from the density gradient energy. 
Actually, when we consider the Thomas-Fermi limit, the gradient energy of 
the vortex core decreases, so that the structure such as Fig.~\ref{Fig1ponch} is expected to 
persist. The example is shown in Fig.~\ref{numericalsolu}(f), where the particle number is 
three times larger than that of Fig.~\ref{numericalsolu}(e). 
In this parameter setting, the domain wall is nearly parallel to the $z=0$ plane 

\begin{figure}
\begin{center}
\includegraphics[width=0.98 \linewidth,keepaspectratio]{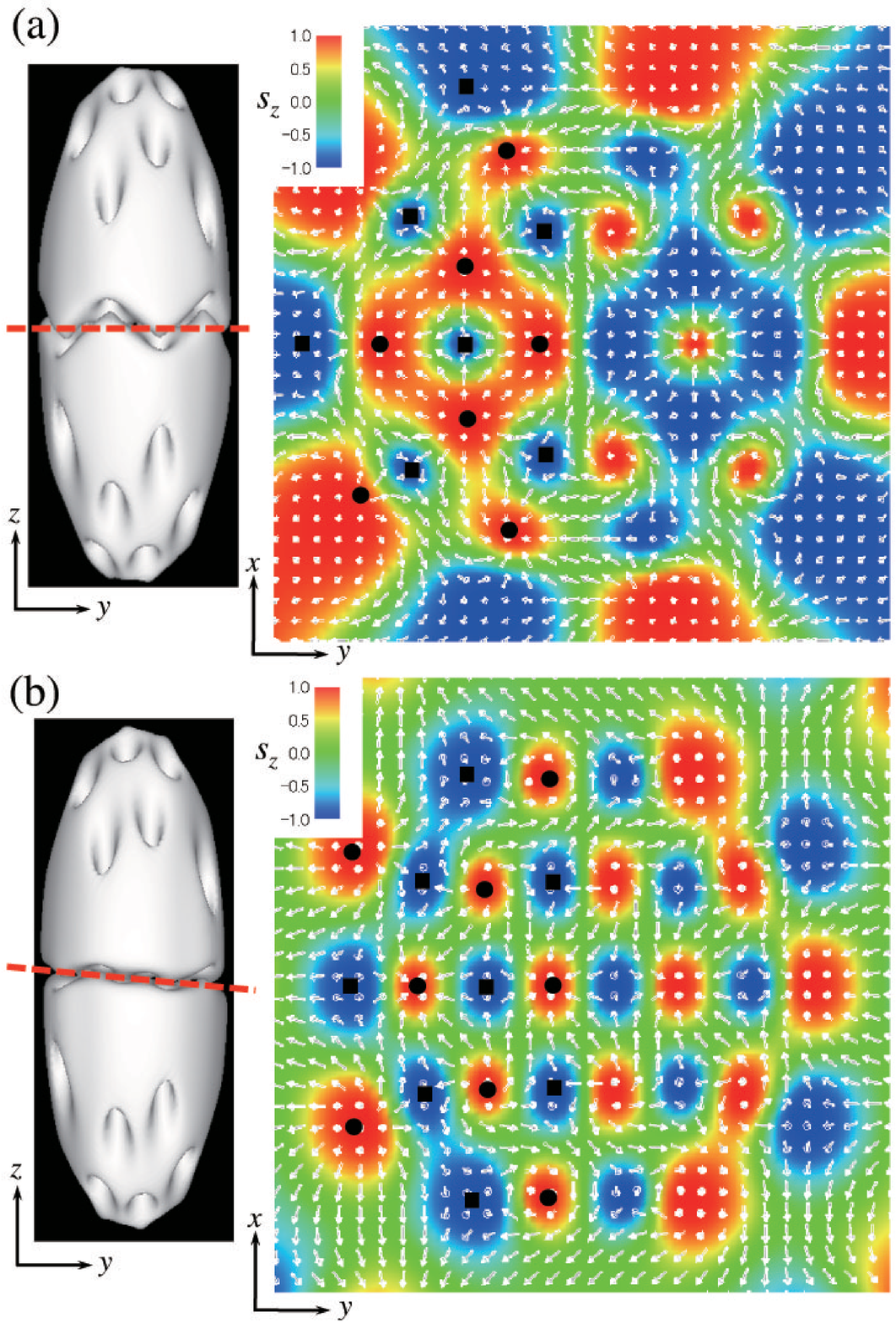}
\end{center}
\caption{(Color Online) The equilibrium structure and 
the spin texture along the $\langle s_{z} \rangle = 0 $ plane for $\tilde{\Omega} = 0.8$, 
$\lambda = 1/4$, $N=1.65 \times 10^5$, and (a) $a_{12} = 2 a$ and 
(b) $a_{12} = 4 a$; the parameters in (a) are the same with those 
of Fig.~\ref{numericalsolu}(f). Left panels show isosurface of the density difference 
$|\rho_1-\rho_2|$ as in Fig.~\ref{numericalsolu}. Right panels show 
the distribution of the pseudospin field ${\bf s}({\bf r})$ in the plane (a) along the $z=0$ plane 
and (b) slightly declined from $z=0$ plane, indicated by the bold lines in the left panels. 
The color scale shows the magnitude of $s_z$. The circles and squares mark the position of 
$\Psi_2$- and $\Psi_1$-vortices, respectively, where we only mark in the $y<0$ region for clarity. }
\label{fig:2}
\end{figure}

Another interesting property of this system at high rotation frequencies is that  
an ordering structure of many interface defects can emerge due to the complicated 
interaction effect. In each domain far from the domain wall, 
singular vortices form a Abrikosov triangular lattice. 
However, singular vortices become coreless vortices near the domain wall and 
a lattice of 2D skyrmions forms on the domain wall. 
Typical example is shown in Fig.~\ref{fig:2}. 
It is important to notice that $\rho_1= \rho_2$ on the domain wall and  
a miscible state is effectively realized in this restricted 2D system of the immiscible condensates. 
In this effective 2D system, the intercomponent coupling $g_{12}$ may be modified 
as $g_{12}^{\rm eff}$, which determines the lattice structure of the 2D skyrmions. 
Note that a lattice of 2D skyrmions prefers a square lattice to a triangular lattice \cite{Mueller}. 
In the parameter setting in Fig.~\ref{fig:2}, the vortex endpoints are shifted relative to each other 
on the domain wall to form a rectangular lattice of 2D skyrmions. 
With increasing $g_{12}$, we can see that the tendency to form a rectangular lattice 
from a triangular lattice becomes remarkable by comparing (a) and (b).
These features are consistent with the phase diagram of a vortex lattice in 2D miscible 
two-component BECs \cite{KTUreview,Mason,Mueller}, 
and may originate from the static vortex-vortex interaction \cite{Aftalionpeak}, 
which is absent in the BPS solution. 
Our numerical solutions show that the domain wall in Fig.~\ref{fig:2}(b) is inclined 
from the $z=0$ plane in order to elongate shorter sides of the rectangles. 
This suggests that the inclination is not accidental but caused by the energetic constraint 
to realize a square lattice. 

\section{Discussion and conclusion}\label{concle}
We have shown that a wall-vortex composite soliton, referred to 
as a D-brane soliton in field theoretical models, 
can be realized as an energetically stable solitonic object in 
phase-separated rotating two-component BECs. Based on the NL$\sigma$M 
derived from the two-component GP model, 
we obtain the analytic solution of the topological solitons, such as domain walls, vortices, 
and their complexes, by taking the BPS bound of the total energy, 
which is a widely used technique in the field theory \cite{Manton}. 
The topological solitons in trapped BECs are found to have the almost same character with 
the BPS saturated soliution in NL$\sigma$M. 
The inhomogeneity of the total density modifies the profile of the soliton quantitatively 
through the reduction of the domain wall tension. 
The domain wall pulled by a vortex is logarithmically bent as the BPS wall in 
the NL$\sigma$M, but it bends more flexibly than expected by 
the tension of the BEC domain wall. 
The numerical analysis of full 3D simulations reveals that the complicated energetic 
constraint has an influence in determining the equilibrium configuration, such as 
the surface tension of the wall, the gradient energy of the density, and interactions 
between vortices and those between interface defects. 
The last statement opens a problem how to consider the properties of an effective 
2D system realized in an interface of multicomponent condensates, which can be 
affected by the extra dimensions (bulk regions). 

It should be noted, however, that there is one significant difference of the wall-vortex composite soliton 
between BECs and the NL$\sigma$M. In the BECs described by the GP model, the total density 
vanishes in the singular vortex core for $n_2 =1$ because the density of nonrotating
component does not enter into the vortex core, as seen in Fig.~\ref{Sigmafitting2}(c). 
The coreless vortex near the domain wall shrinks 
to a singular vortex for a finite distance and thus we can identify a point 
connecting a singular vortex and a coreless vortex. This
is in contrast to the case in the NL$\sigma$M, where a coreless vortex extends to infinity along the 
thin vortex core, avoiding the singularity since ${\bf s}$ is well-defined everywhere. 
Hence, a connecting point is absent, or more precisely, it should be positioned at infinity in this model. 
In a field theoretical model, the connecting point forms defect called ``boojum", which serves the negative binding
energy of vortices and a wall and a half of the negative
charge of a single monopole \cite{Isozumi,Sakai}. 
Boojums are known as point defects existing upon the surface of the ordered phase; the name was 
first introduced to physics by Mermin 
in the context of superfluid $^3$He \cite{Mermin}. 
Boojums can exist in different physical systems, such as the interface separating A and B phases 
of superfluid $^3$He \cite{Blaauwgeers,Volovik}, 
liquid crystals \cite{Kleman}, the Langmuir monolayers at air-water interfaces \cite{Fischer}, 
multi-component BECs with a spatially tuned interspecies interaction \cite{Takeuchi,Borgh}, 
and high density quark matter \cite{Cipriani}.
In the present model, boojums can be found at the end points of vortices on the domain wall, at which 
the vortices change their character from singular to coreless type. 
A suitable topological charge for boojums in two-component BECs can be derived 
by noting the analogy of the Abelian gauge theory \cite{Kasamatsuissue}. 
A detailed study of the distribution of the boojum charge and the interactions between 
boojums remains as a future study.

As pointed out in Ref. \cite{KasamatsuD}, the domain wall in two-component BECs are 
useful to simulate some analogue phenomena of the D-brane physics in a laboratory. 
One famous example is a nonequilibrium dynamics such as brane-antibrane annihilation, 
which was proposed for a possible explanation of inflationary universe in string theory. 
In braneworld scenarios of cosmic inflation the annihilation may lead to defect production 
that could be directly observed in atomic BECs;
the experiment has been performed with superfluid $^3$He A-B interfaces, but the detection 
of defects is difficult \cite{Bradley}.
Recently, we proposed that domain-wall annihilation in two-component BECs actually 
demonstrates a brane-antibrane collision and a subsequent creation of cosmic strings, 
causing {\it tachyon condensation} accompanied by spontaneous Z$_2$ symmetry breaking 
in a two-dimensional subspace \cite{Takeuchitac}. 
Also we propose that, when strings are stretched between the brane and the antibrane, namely when 
the filling component has vortices perpendicular to the wall, ``cosmic vortons" can emerge via 
the similar instability \cite{Nitta}. 
All of these phenomena can be monitored directly in experiments. 
We hope that our works open a 
new trend of the cold atom physics as ``simulator of everything". 

\begin{acknowledgments}
This work was supported by KAKENHI
from JSPS (Grant Nos. 21340104, 21740267
and 23740198). This work was also supported by the
``Topological Quantum Phenomena" (Nos. 22103003 and
23103515) Grant-in Aid for Scientific Research on Innovative
Areas from the Ministry of Education, Culture,
Sports, Science and Technology (MEXT) of Japan.

\end{acknowledgments}



\begin{thebibliography}{99}
\bibitem{tyuu}
Even in the present scheme, where the same rotation is applied for 
the both components, the stabilization of a vortex only in the one component 
is possible if we choose the suitable asymmetric seed for two components 
in the initial condition because of the hysterisis effect \cite{KasamatsuD,GarciaRipoll}; 
the critical rotation frequency for vortex nucleation and that for vortex stabilization 
is generally different. 
Preparing vortices only in one-component is attainable more easily when 
there is an imbalance of the parameters for $\Psi_{1}$- and $\Psi_{2}$-component.
\end{thebibliography}
\end{document}